\documentclass[fleqn,10pt]{wlscirep}
\usepackage[utf8]{inputenc}
\usepackage[T1]{fontenc}

\usepackage{hyperref}
\usepackage{amsmath,graphicx,makecell}
\usepackage{enumitem}
\graphicspath{{./fig/}}
\usepackage{threeparttable}
\usepackage{booktabs}
\usepackage{algorithmic}
\usepackage{subcaption}
\usepackage[linesnumbered, ruled, vlined]{algorithm2e}
\usepackage{todonotes}   
\usepackage{soul} 
\newcommand{\NA}{---}

\usepackage[markup=underlined]{changes}
\definechangesauthor[color=blue]{Author}
\usepackage{todonotes}
\setcommentmarkup{\todo[color={authorcolor!20},size=\scriptsize]{#3: #1}}

\title{An Open-source Extendable Model and Corrective Measure Assessment of the 2021 Texas Power Outage}

\author[1,$\dagger$]{Dongqi~Wu}
\author[1,$\dagger$]{Xiangtian~Zheng}
\author[2]{Yixing~Xu}
\author[2]{Daniel~Olsen}
\author[2]{Bainan~Xia}
\author[1]{Chanan~Singh}
\author[1,3,*]{Le~Xie}

\affil[1]{Department of Electrical and Computer Engineering, Texas A\&M University, College Station, Texas, USA}
\affil[2]{Breakthrough Energy Sciences, Seattle, Washington, USA}
\affil[3]{Texas A\&M Energy Institute, College Station, Texas, USA}
\affil[*]{Corresponding author: le.xie@tamu.edu}
\affil[$\dagger$]{Co-first author}

\begin{abstract}
Unprecedented winter storms that hit across Texas in February 2021 have caused at least 69 deaths and 4.5 million   customer interruptions  due to the wide-ranging generation capacity outage and record-breaking electricity demand. 
While much remains to be investigated on what, how, and why such wide-spread power outages occurred across Texas, it is imperative for the broader macro energy community to develop insights for policy making based on a coherent electric grid model and data set. In this paper, we collaboratively release an open-source extendable model that is synthetic but nevertheless provides a realistic representation of the actual energy grid, accompanied by open-source cross-domain data sets.
This simplified synthetic model is calibrated to the best of our knowledge based on published data resources. Building upon this open-source synthetic grid model, researchers could quantitatively assess the impact of various policies  on mitigating the impact of such extreme events. As an example, in this paper we critically assess several corrective measures that could have mitigated the blackout under such extreme weather conditions.
We uncover the regional disparity of load shedding. The analysis also quantifies the sensitivity of several corrective measures with respect to mitigating the severity of the power outage, as measured in Energy-not-Served (ENS). This approach and methodology are generalizable for other regions experiencing significant energy portfolio transitions. 
\vspace{-1.8em}
\end{abstract}

\begin{document}
\flushbottom
\maketitle
\thispagestyle{empty}


\section{Introduction}

The extreme winter storm and associated electricity outages in February 2021 are estimated to have caused more than 70 deaths \cite{bk2} and \$200 billion economic loss \cite{bk3} in the state of Texas. Besides the official brief review \cite{ercot_review} and ongoing internal investigation on the Electric Reliability Council of Texas (ERCOT), there have been preliminary reports from non-peer-reviewed  articles \cite{article_1, iea_commentary, gas_scarcity}, press interviews \cite{report_1,report_2} and a few recent academic publication \cite{busby2021cascading} on  potential causes and  technical solutions for this blackout event. 
{A challenging question for the broader energy research community lies in how to develop an open-source and extendable model that captures the key characteristics associated with this extreme outage. Such models would offer an open platform for corrective policy assessment. }

The existing studies of power grid resilience to extreme weather conditions offer power grid resilience enhancement strategies mainly in two perspectives: (a) physical hardiness such as vegetation management \cite{gholami2018toward}, substation relocation \cite{bresch2010building}, selective undergrounding \cite{kopsidas2017power} and etc., and (b) operational capability such as emergency generator \cite{kenward2014blackout}, distributed energy resources \cite{ton2015more, liu2016microgrids}, grid monitoring system \cite{panteli2013assessing, eskandarpour2016machine} and etc. However, the lack of a common open-source realistic power grid model prevents broader communities from the assessment of these methods for specific real-world events via simulation.
While researchers recently have contributed to the creation of large-scale synthetic grid models \cite{birchfield2016grid} for analysis such as macro-scope energy portfolio transition \cite{BTE1,BTE2}, 
cross-domain and open-source reliable approaches to quantify impact from corrective measures against blackout events under extreme frigid weather are still at a nascent stage, with several gaps in existing research. 
First, existing open-source large-scale synthetic grid models are not ready-to-use for the event reproduction without rigorous calibration. Second, the lack of aggregated and processed event timeline data prevents exhaustive simulation and further investigation. Last but not least, the lack of consistent quantified criteria renders studies on the effectiveness of potential corrective measures and their combined effect incomparable.

Here, we collaboratively develop an open-source large-scale synthetic baseline grid \cite{git}, providing a realistic representation of the actual Texas electric grid, accompanied by the open-source data set along the event timeline. This ready-to-use multi-platform synthetic grid model is calibrated based on open-source data sets, including generation by source, load by weather zones, generation unit outage timeline, load shedding record, etc. To the best of our knowledge, it is the first fully open-source approach to model, simulate, benchmark, as well as propose corrective measures against the 2021 Texas power outage. 
{The blackout event reproduction results obtained from this open-source synthetic model are compared and validated using key parameters obtained from the actual blackout reports, including the change of system generation capacity, load demand and load shedding data.}
Additionally, we propose and evaluate multiple technical solutions that can possibly mitigate the electricity scarcity under such extreme weather conditions, including energy system winterization, interconnected HVDC lines, up-scaled demand response program and strategic energy storage facilities. Leveraging the synthetic grid, we perform quantitative analysis on the corrective measures in the aspect of reducing the extent of blackout events. Our results indicate the strong disparity among the winterization effectiveness for generation units of various source types and geographical regions, the quantitative assessment of certain corrective portfolios, and the interdependence of per-unit performance of corrective measures.

\section{Open-source Synthetic Grid Model and Data}
We first collaboratively develop an open-source, large-scale, synthetic baseline grid that provides a realistic representation of the actual Texas electric grid, and then integrate generation and load-related data along the event timeline, which are both publicly available on Github \cite{git}. The original sources are detailed in the documentations in the Github repository (see Data and Code Availability). In this paper, the synthetic model creation  focuses on feasibility of the direct current optimal power flow (DCOPF) solution without transient stability assessment for the following reasons. 
{First, this DCOPF-based synthetic grid model offers substantial insights to the chief parameters associated with this blackout event based on the public data availability. Introducing synthetic yet unrealistic dynamic transient parameters would be counter-productive for the purpose of open-domain analysis and corrective measure assessment. }
{Second, this model is extendable further for the research community if more credible dynamic and detailed parameters become available.}
{Third, this model is shown to be computationally efficient. It can serve as a bridge to connect the macro energy systems research community with the electric power systems engineering community. }


For the purpose of `what-if' analysis, we create a comprehensive blackout event dataset via collection from publicly available sources \cite{ercot_gridinfo, ercot_review, ercot_generator_outage2, eia_capacity,poweroutage} and estimation (see Methods). This dataset integrates actual load by weather zone, actual generation by source, 7-day-ahead load forecast by source, solar and wind generation forecast, generation units outage, actual available generation capacity, actual load shedding and customer power outage into a single ready-to-use format. Here, we define the \textbf{\textit{counterfactual load}} as the 7-day-ahead load forecast
and the simulated \textbf{\textit{load shedding}} as the gap between the post-shed and counterfactual load, 
and introduce the \textbf{\textit{estimated generation capacity}} by weather zone based on rated generation capacity, thermal generation units outage and actual renewable generation (see Methods), all of which play important roles in the what-if analysis. 

In this paper, we develop an open-source synthetic grid that captures some of the key characteristics of the Texas Interconnection. Texas Interconnection is one of the AC synchronized grids in North America that covers most of Texas. The Texas Interconnection has a total of more than 86,000 MW of generation resources of various types, including 51.0\% Natural Gas, 24.8\% wind, 13.4\% coal, 4.9\% nuclear and 3.8\% solar \cite{ercot_factsheet}. The loads in the Texas Interconnection are further divided into 8 weather zones and the various grid operation and market data are aggregated to the zone level before publishing. There are two DC ties between the Texas Interconnection and the Eastern Interconnection that allow power exchange between two un-synchronized power grids.

In this paper, the synthetic grid is calibrated carefully to capture the key open-domain statistical characteristics of the Texas Interconnection.
The synthetic Texas grid model is adapted from an existing test system \cite{breakthrough_data} and rigorously calibrated in several aspects, namely generator units capacity, and transmission line rating (see Methods). The geographical load distribution comes from the existing grid model \cite{birchfield2016grid}, while their real-time magnitudes in simulation are adjusted according to the real load dataset or calculated ones depending on whether the real load data are available. The generator units capacities are updated to the actual available generation capacity \cite{eia_capacity} in January 2021. Without modifying the network topology, some transmission lines are upgraded to ensure that the model remains feasible in the period leading up to the blackouts and that no renewable generators are unreasonably curtailed due to congestion. 

\begin{figure}[ph]
	\centering
	\includegraphics[width=0.95\textwidth]{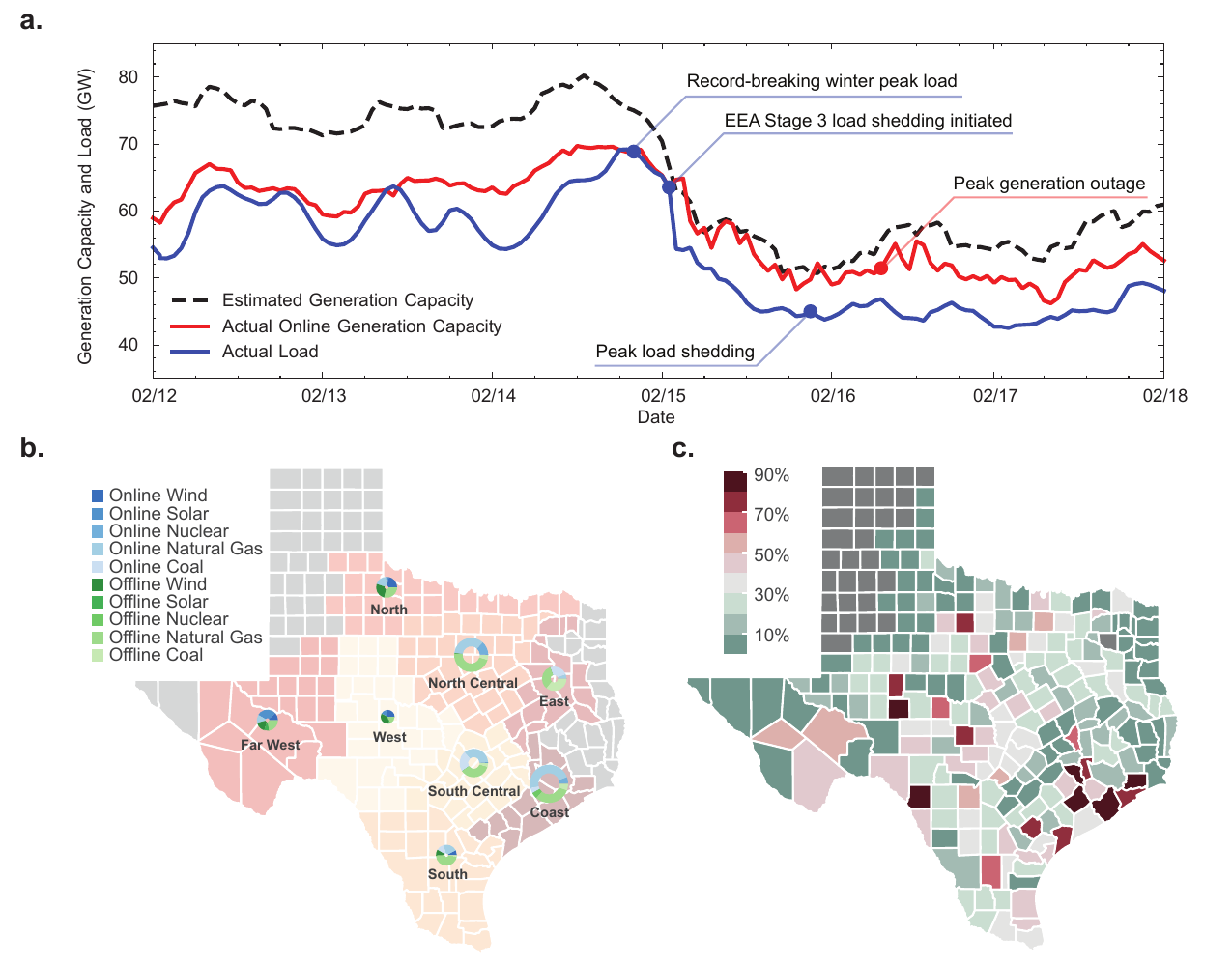}
	\caption{\textbf{Overview of the 2021 Texas blackout event from the perspective of generation capacity and load shedding.} \textbf{a}, Blackout event overview in terms of generation capacity and actual load associated with the key event labels. The significant gap between the estimated generation capacity and actual online capacity before February 15 and increasing mismatch between them after noon on February 16 are in line with expectations due to several reasons explained in Appendix A.1. \textbf{b}, Online and offline generation capacity by source in ERCOT weather zones at 11:00 a.m. February 16. The ring size is determined by the zone-level total generation capacity and its color represents the type of generation capacity. The block color represents the weather zone where the county is located (grey blocks are not within eight weather zones). \textbf{c}, Normalized county-level customer outage percentage during the "darkest" period. The block color represents the county-level customer outage percentage \cite{poweroutage}, where the ERCOT average outage percentage is 31\% (grey blocks mean no data available). The "darkest" period is ranging from 8 p.m. February 15 to 11 a.m. February 16, possessing the largest load shedding amount.  
	}
	\label{event_overview}
\end{figure}

\begin{figure}[ht]
	\centering
	\includegraphics[width=0.5\textwidth]{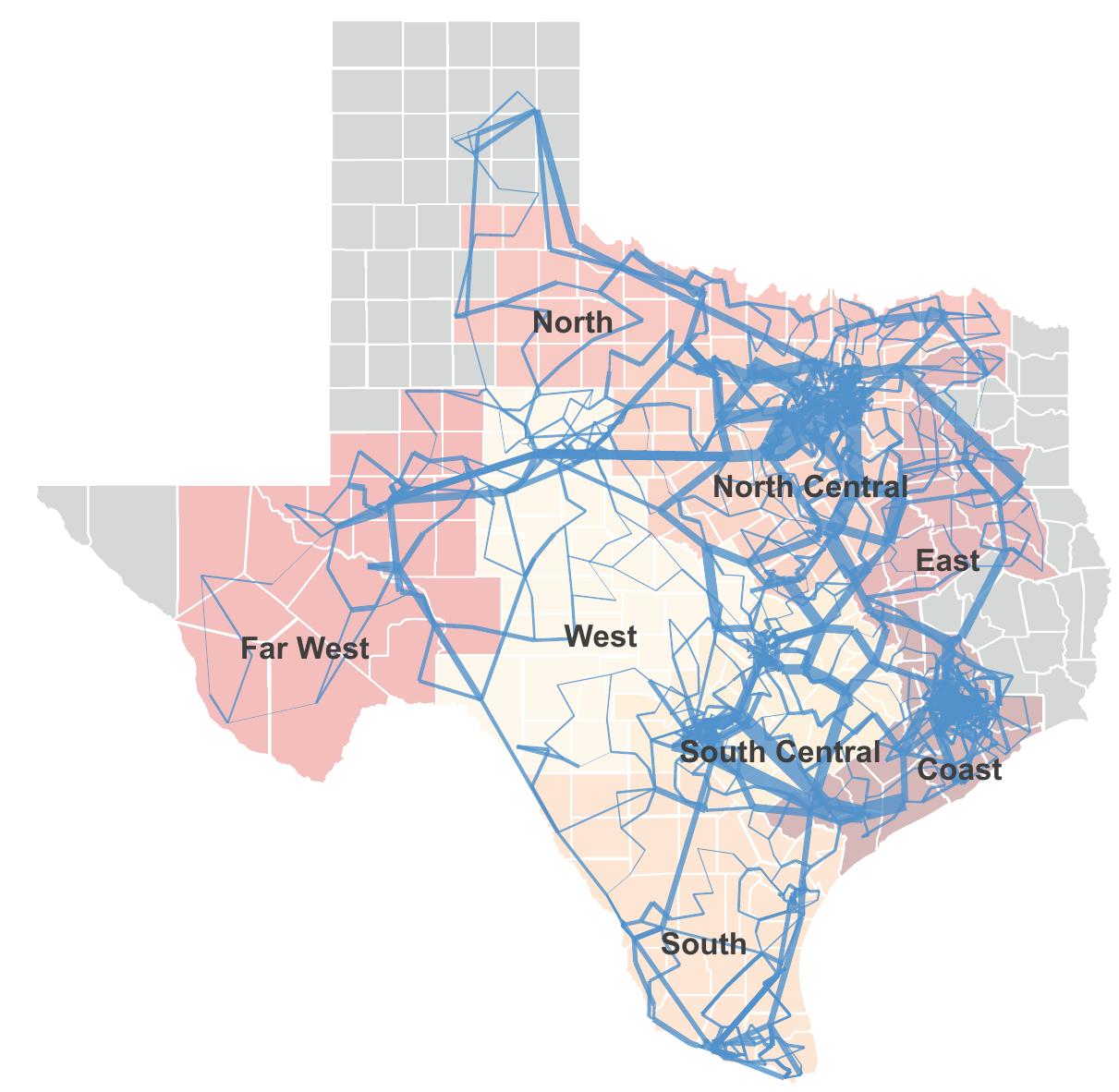}
	\caption{\textbf{Visualization of the large-scale synthetic Texas grid.} Here the branch width is proportional to the transmission line rating. This synthetic grid contains 606 generators, 1,350 loads and 3,206 branches, of which the generation units capacity, transmission line rating and load distribution are calibrated based on the available Texas grid-related data.
	}
	\label{synthetic_grid}
\end{figure}

Integrating the open-source datasets, the ready-to-use synthetic grid model is the first open-source simulation package dedicated to potentially provide firm interdisciplinary insights into the particular real-world blackout event, which is extendable for the broader macro energy community due to its transparency. 
To give an intuitive impression on the blackout event, we provide an event overview along with regional generation outages and customer power outages (Fig. \ref{event_overview}). The timeline of the whole blackout event (Fig. \ref{event_overview}-a) that contains the actual total load, actual and estimated generation capacity shows the electricity scarcity due to the high load demand and wide-ranging generation outage. 
The actual generation and generation outage across eight ERCOT weather zones (Fig. \ref{event_overview}-b) show that the generation outage at the darkest hour mainly consists of natural gas thermal generation outages across ERCOT and renewable generation outages in the North, West, Far West and South zones. We also find the regional disparity of load shedding (Fig. \ref{event_overview}-c) from the aggregated county-level utility-reported customer outage data \cite{poweroutage} during the "darkest" period, namely from 8 p.m. February 15 to 11 a.m. February 16. Specifically, the satellite counties around Houston in the Coast zone and several counties distributed in the West zone suffered the most severe outages. 
We notice the significant gap between the estimated generation capacity and actual online capacity before February 15 and increasing mismatch between them after noon on February 16 that are in line with expectations due to several reasons explained in Appendix A.1. 
We have observed that there exists a substantial mismatch between actual load and either actual online or estimated generation capacity, which is beyond the reserve limit. This mismatch may be attributed to multiple reasons, such as transient stability requirements, reactive power demands, and capped wholesale market price \cite{high_price}, which deserve more investigation but are nevertheless outside the scope of this paper. To show the complex but realistic features of the synthetic grid, we visualize the topology of the whole synthetic grid (Fig. \ref{synthetic_grid}), of which load distribution, generation units capacity and transmission lines rating are calibrated based on the static Texas grid-related data (see Methods). In the following analysis, we will leverage the synthetic model along with the blackout event dataset to demonstrate its fidelity by reproducing the blackout event via simulation and then perform quantitative assessments of multiple corrective measures against extreme frigid weather.

We would like to remark that the approach of developing such an open-source synthetic grid model is extendable for similar assessment of impact and corrective measures of severe weather-induced power outages across many regions.  
 The synthetic model follows the \text{Matpower}  case format which uses tables to store generator, load and topology information \cite{matpower}. The modification of all components are done through altering the corresponding row of the tables. As an example, this synthetic grid model is also compatible with the open-source model and dataset in  \cite{breakthrough_model} which can be used to perform many more comprehensive studies such as HVDC interconnection designs, 2030 projected profiles, renewable and storage scenarios and integrated simulation along with the synthetic Western and Eastern Interconnections.

\section{Methods}

\subsection{Data Aggregation}
In order to validate the model by event production and perform quantitative what-if analysis, we integrate the blackout related data during the event period between February 14 to February 18. The original sources for all datasets are provided in the Data and Code Availability section. We integrate the datasets via two ways, namely data collection from multiple resources \cite{ercot_gridinfo,eia_capacity,ercot_generator_outage2,poweroutage} and data estimation.

\begin{figure}[h]
	\centering
	\includegraphics[width=1\textwidth]{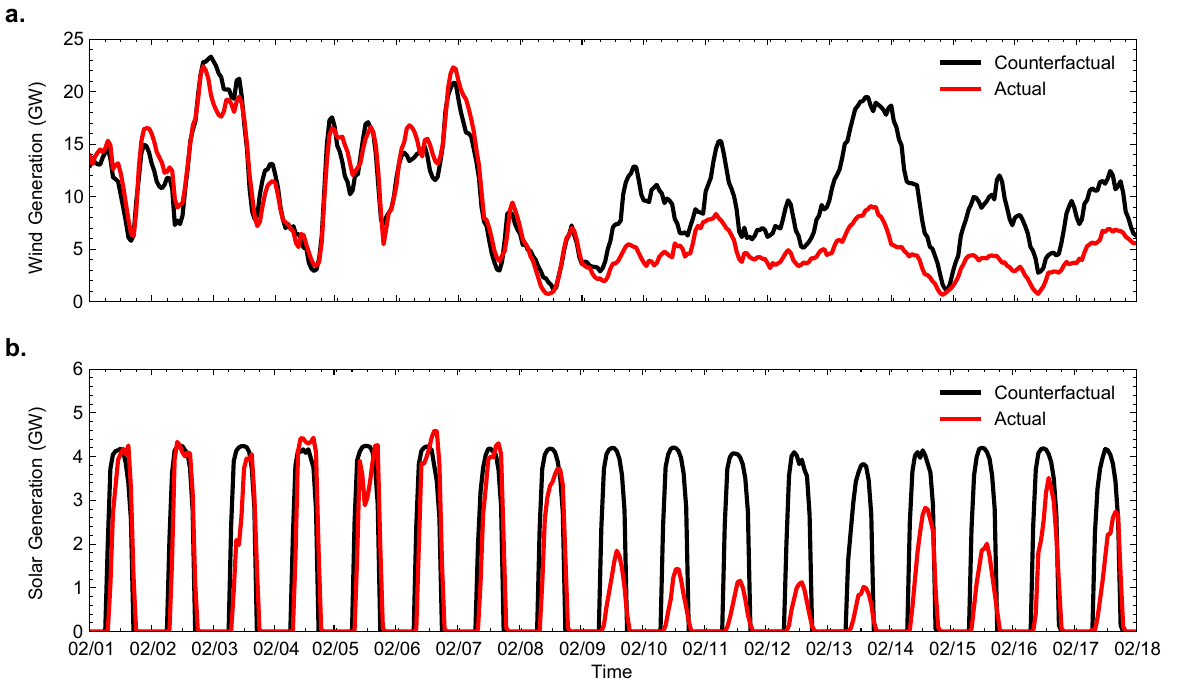}
	\caption{\textbf{Actual and counterfactual wind and solar generation.} Note that the counterfactual wind generation is estimated based on the weather data \cite{wind_weather_data} as described in the literature \cite{BTE1}, while the counterfactual solar generation is obtained from ERCOT solar forecast data \cite{ercot_gridinfo}. The model-based renewable generation estimation is preferred because of its higher granularity. However, for the case of counterfactual solar generation, the lack of relevant weather data prevents this model-based method and instead we have to use the ERCOT solar forecast data.
	}
	\label{solar-and-wind}
\end{figure}

\begin{itemize}
    \item \emph{Data Collection}: We collect actual load, actual generation, 7-day-ahead load forecast and 7-day-ahead solar generation forecast data from the ERCOT regular data channel \cite{ercot_gridinfo}. 
    We collect generation units outage data from the source \cite{ercot_generator_outage2} dedicated for the blackout event, which specifies the outage period, outaged capacity, source type, and location of outaged and de-rated generator units. Note that only part of all generation outages are included in this source, since some resource entities do not provide ERCOT consent to disclose, and outages shorter than two hours may not be included.
    We obtain generation capacity data from the Energy Information Admission (EIA) generation inventory data source \cite{eia_capacity}.
    Additionally, we get the customer power outage data \cite{poweroutage} from PowerOutage, which has city-level utility-reported number of customers suffering power outage.
    \item \emph{Data Estimation}: We define the estimated generation capacity as the sum of total maximum online capacity of thermal and nuclear generation and total real-time varying available wind and solar generation capacity. Here the maximum online capacity of thermal and nuclear generation are equal to the seasonal maximum capacity \cite{eia_capacity} subtracted by the generation units outage \cite{ercot_generator_outage2}, while the real-time varying capacity of wind and solar generation is the aforementioned collected wind and solar generation data \cite{ercot_gridinfo}. We define the load shedding as the gap between the actual and counterfactual load data. Here the counterfactual load data refer to the 7-day-ahead load forecast. We also estimate the counterfactual wind generation as described in the literature \cite{BTE1} using the associated weather data \cite{wind_weather_data} during the blackout event period. Wind generation estimation in this way can achieve the highest granularity. Due to the lack of weather data required by solar generation estimation model, the 7-day-ahead forecast solar generation is the alternative way for the counterfactual estimation. Great matches between actual and counterfactual wind and solar generation profiles before February 9 as shown in Fig. \ref{solar-and-wind} demonstrates the accuracy of counterfactual generation, and it also indicates that the impacts of the winter storm on renewable generation approximately started from February 9.
\end{itemize}
\subsection{Synthetic Grid Creation}

The synthetic Texas grid model is adapted from the existing test system \cite{breakthrough_data}, and rigorously calibrated in several aspects, namely, system topology, geographical load distribution, generation units capacity and transmission lines rating. Note that although the latest released synthetic Texas 7,000-bus grid\cite{texas_7000} offers more granularity in terms of buses and zones, the zones do not line up with the data available from ERCOT, rendering remarkable difficulty for model calibration. Therefore, we still choose this 2,000-bus case as the base model.
\begin{itemize}
    \item \emph{Load distribution}: The geographical load distribution comes from the original design of the synthetic grid model \cite{birchfield2016grid}, of which the relative sizes are reasonably determined by the demographic and geographical information. The real-time sizes of loads in simulation are further adjusted to the real hourly load dataset or calculated counterfactual ones. As the public datasets only disclose aggregated load of each weather zone, all loads within each weather zone are proportionally scaled such that the sum of all loads in each area equals the published numbers. 
    \item \emph{Generation units capacity}: The existing test system contains generation capacity inventory up to 2016 \cite{birchfield2016grid} plus the largest generators added in the period from 2017 to 2019, with the entire generator fleet scaled to match totals by type and zone at the start of 2020 \cite{breakthrough_data}. To update this dataset to 2021 conditions, we add eight new generators: two natural gas generators (a total of 427 MW), one biomass generator (100 MW), four solar generators (742 MW), and one wind generator (220 MW). These generators were added at high-voltage buses in the synthetic network closest to their real locations. Finally, the generation fleet was scaled such that totals by type and weather zone match the totals from EIA Form 860-M, December 2020 (to account for any uncaptured additions, retirements, de-ratings, etc.).
    \item \emph{Transmission lines rating:} The topology of the system is adopted from the synthetic Texas 2,000-bus grid as described in the literature \cite{birchfield2016grid}. Without modifying the topology of the transmission network, their capacity ratings and imdepances are tuned to emulate transmission expansion to accommodate the additional generation resources, additional HVDC interconnections and 2021 demand. Transmission lines are upgraded as necessary to ensure network feasibility and to avoid unrealistic curtailment of variable renewable generators. Specifically, 59 out of 3,206 branches are upgraded, representing 108.4 GW-miles of transmission lines upgrades (out of a total of 19,374 GW-miles in the base grid. These transmission upgrades are mostly in the Far West (29) and North (14) weather zones, where growth of demand and generation resources has been greatest relative to the 2016 transmission network capacities.
\end{itemize}

\subsection{Load Shedding and Restoration Operation Principles}
In ERCOT operation protocols, when the system-wide reserve drops to a dangerous level that qualifies for Energy Emergency Alert (EEA) conditions, the grid operator will use different resources from various participants of the ERCOT market to maintain grid security. In EEA levels 1 and 2, ERCOT will first contact industrial loads that agreed to be disconnected during emergencies and call upon available demand response programs. In EEA level 3 events during which the operating reserve capacity is below 1000 MW, ERCOT will ask transmission companies to shed load, typically done through rotating outages. In our simulations, we aim to follow a similar process in determining the total amount of load to shed while maintaining simplicity and generality. For each snapshot, we start by applying the counterfactual load and try to find a feasible power flow solution. If the available capacity cannot satisfy the full demand or the supply is limited by transmission line congestion, DCOPF will be infeasible. In that case, we then gradually reduce the load across the network until a feasible solution is found. In the reproduction simulation, the spatial distribution of shedded load obtained from the ERCOT historic demand data during the event is adopted into the simulation to more closely mimic the timeline of the event. Similarly, if the system-wide reserve is high enough and there is active load shedding from the past hours, we attempt to slowly reconnect them back until the operation reserve has been depleted. The full logic flow of load shed in simulation is listed in Algorithm \ref{algo:ls}.
    
\begin{algorithm}
\caption{Iterative Load Shedding in Simulation}
\label{algo:ls}
\begin{algorithmic}
\STATE Load renewable generation profile of hour $t$, $P_{g}^t$ into model
\STATE Modify thermal generator capacity based on unit outage data
\STATE Load counterfactual load of hour $t$, $ P_{l}^t$ into model
\STATE Apply load shedding from the past hour $P_{ls}^{t-1}$ to load buses
\STATE Compute system-wide capacity reserve $P_{r}^t = \sum_i (P_{g,i}^t - P_{l,i}^t) + P_{ls}^{t-1}$ $\forall i \in $ [list of all buses]
\STATE Attempt to solve DCOPF on the base profile
\IF{DCOPF is infeasible  or $P_{r}^t < P_{r,min}$}
\WHILE{DCOPF is infeasible or $P_{r}^t < P_{r,min}$}
\STATE Increase load shedding: $P_{ls}^t = P_{ls}^t + \Delta P_{ls}$
\STATE Update $P_{r}^t = P_{r}^t + \Delta P_{ls}$
\ENDWHILE
\ELSIF{Operation reserve $P_{r}^t > P_{r,min}$ and $P_{ls}^{t} > 0$}
\WHILE{OPF is feasible and $P_{r}^t > P_{r,min}$ and $P_{ls}^t > 0$}
\STATE Decrease load shedding: $P_{ls}^t = P_{ls}^t - \Delta P_{ls}$
\STATE Update $P_{r}^t = P_{r}^t - \Delta P_{ls}$
\ENDWHILE
\ENDIF
\STATE Save $P_{ls}^t$ as the minimum load shedding for hour $t$
\end{algorithmic}
\end{algorithm}
    
 \begin{figure}[ht]
	\centering
	\includegraphics[width=1\textwidth]{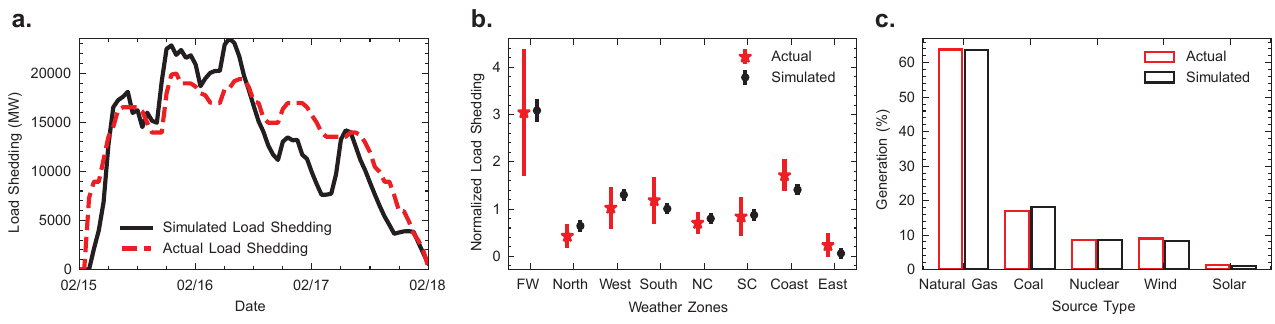}
	\caption{\textbf{Blackout event reproduction via simulation on the synthetic grid in comparison with the real data.} \textbf{a}, Comparison between the simulated and actual total load shedding curves. \textbf{b}, Zone-level normalized load shedding with 95\% confidence interval (CI) during the period between 8 p.m. February 15 and 12 p.m. February 16 that has the highest load shedding. FW, NC and SC respectively represent Far West, North Central and South Central zones. Here the normalized load shedding refers to the ratio of load shedding percentage over online load percentage for each zone. We use the estimated load shedding data based on the counterfactual and actual load data from ERCOT rather than customer outage data \cite{poweroutage} as the real benchmark (See Appendix A.3). \textbf{c}, Comparison between actual and simulated percentage of generation compositions of various source types throughout the event.
	}
	\label{reproduction_result}
\end{figure}

\section{Model-based Simulation of the 2021 Texas Blackout Event}
{To validate the calibrated model dedicated for the blackout event, we reproduce the Texas blackout event from February 15 to February 18. To this end,} we simulate the synthetic grid model using the aggregated data, where realistic load shedding allocation and DCOPF are key steps. We take the \textbf{\textit{estimated generation capacity}} (the binding constraint for load shedding) and \textbf{\textit{counterfactual load}} (the ebb-flow pattern of load) as the simulation inputs. To achieve the fidelity of load shedding, we mimic the guides of load shedding and restoration  \cite{nodal_operation_guides, ercot_training_manual} to determine the total load shedding amount at any given moment, and perform appropriate spatial allocation of load shedding to reflect its regional disparity (see Methods). We reproduce the blackout event by iteratively solving DCOPF given the post-shedding load (see Methods), which reveals the hourly change of geographically distributed load, generation and load shedding across Texas in detail.

 We demonstrate the fidelity of the synthetic grid and the associated simulation methods by the reproduction results of load shedding and generation composition (Fig. \ref{reproduction_result}). To quantify the severity of the blackout event, we use the power system reliability index \textbf{\textit{energy-not-served}} \cite{singh2018electric} (ENS), defined as the integral of load shedding over the event timeline, to quantitatively evaluate the load shedding throughout the rest of this paper. We first demonstrate the fidelity of the geographical load distribution and the designed load shedding algorithm by the good match between the actual and simulated total load shedding (Fig. \ref{reproduction_result}-a) that respectively represent a total of $998.8$ GWh and $929.6$ GWh Energy-Not-Served {(6.91\% difference). The correlation coefficient between the two timeseries is 0.88, which also indicates a good match between the re-produced load shedding process and actual ERCOT record.} The unavoidable mismatch attributes to the combined effects of errors in synthetic grid modelling and system operation under emergency conditions (see Appendix A.2 for the remark on the mismatch).
 We then validate that the simulation well captures the regional disparity of load shedding across eight weather zones \cite{weather_zone_map} by comparing the simulated zone-level normalized load shedding with the real one (Fig. \ref{reproduction_result}-b). 
It shows that Far West experienced the most disproportional load shedding among all zones and Coast has suffered from a significantly worse condition compared to the other two most populous zones: North Central and South Central. 
Finally, we validate the fidelity of generation units capacity by type and generation cost curves used for DCOPF by showing the almost perfect match between actual and simulated generation composition throughout the event (Fig. \ref{reproduction_result}-c). 
The reproduction results validate the synthetic baseline model, the related data and the associated simulation methods, which provide a reliable basis for the following what-if analysis. Additionally, the reproduction results can transparently show the change of load, generation and load shedding along the timeline, aiding public multidisciplinary researchers in combing the event development process, investigating the event causes and providing possible technical solutions. The transparency and reproducibility of the synthetic grid model also allow public researchers to contribute to further model development and calibration.
 

\subsection{Details of the Event Simulation}
\begin{itemize}
    \item \emph{Load and Load Forecast Profile}:
    In the reproduction of the outage event, we apply the real historic load during the period between February 12 and February 18 to the synthetic Texas network. The ERCOT load data are divided into eight weather zones in Texas, each containing a specific set of counties. In the simulation, we scale the base value of every load bus within each weather zone such that the total load capacity in the zone equals the ERCOT load forecast data on the same hour, which is used as the counterfactual load profile.
    
    \item \emph{Renewable Generator Capacity}:
    We have estimated the unit-level wind and solar capacity data during the event periods using the combination of available weather data and ERCOT generation-mix data. In the reproduction of the outage event, the unit-level data are all scaled according to the total actual renewable generation data published on EIA \cite{eia_capacity}. These data are set to be the maximum real power output for renewable generators and is subject to further curtailments, should congestion in the lines occur. The renewable generators are set to have zero cost in the DCOPF formulation to prioritize them over thermal energy sources.
    
    \item \emph{Dispatch-able Generator Capacity}:
    We use the ERCOT unit-level outage report\cite{ercot_generator_outage2} to determine the maximum real power output capacity for thermal generators. Although the maximum rate capacity for thermal plants is fixed through the event, many thermal generators have experienced outage or de-rating due to various reasons, including limited fuel supply, facility freezing and planned maintenance. As the generators in the synthetic networks are equivalent generators, exacting matching with the outage report is not possible. Instead, the outage data are first aggregated to county-level and used to set the maximum capacity of all thermal generators of the same fuel type in each county.
    
    \item \emph{Formulation of DCOPF}:
    Using deterministic system demand and renewable generation data, the actual output of dispatchable generators (coal, natural gas and nuclear) and the power flow pattern in the network is determined by Direct-Current Optimal Power Flow (DCOPF). DCOPF is an optimization formulation that computes the most economic real power output assignment for all dispatch-able generators in a network, such that all transmission line flow is below their thermal limit and the total real power generation equals the total demand across the network. The Matlab package Matpower is used to solve DCOPF for all simulations. A detailed description and formulation of DCOPF implemented by Matpower can be found in their documentation \cite{mpopf}.
    
    \item \emph{Cost Curve}: The cost curves of generators in the original synthetic network\cite{birchfield2016grid} are revised to make an even closer match with reality. We have run a year-round multi-period DCOPF simulation using 2016 ERCOT renewable and load profile data to confirm that the total generation from each of the fuel sources is sufficiently closely matching historical data. A detailed experiment procedure and result comparison can be found in the literature \cite{BTE1}.

\end{itemize}

\section{Quantitative Assessment of Corrective Measures against Extreme Frigid Weather}

\begin{figure}[h]
	\centering
	\includegraphics[width=0.6\textwidth]{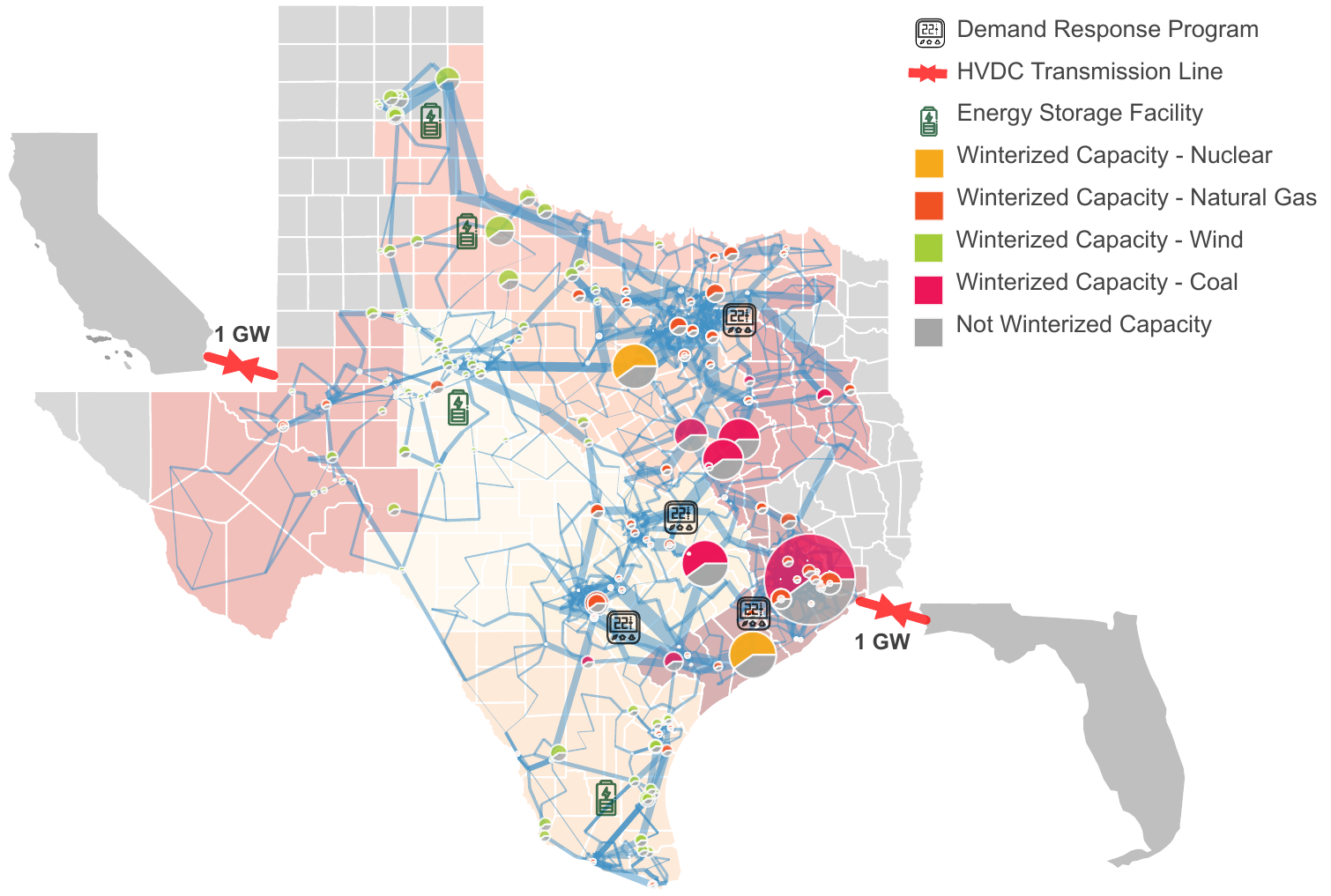}
	\caption{\textbf{Conceptual diagram of geographical distribution of four corrective measures in simulation.} It illustrate a combination of 60\% facility winterization, 2 GW HVDC lines, 2 GW up-scaled demand response program and 4 GWh strategic energy storage facility. Here the generators of thermal and renewable energy evenly implement 60\% winterization. The two long-distance HVDC lines with total 2 GW deliver the power from California through the Western Interconnection to the West zone and from Florida through the Eastern Interconnection to the Coast zone. The up-scaled demand response program is mainly deployed in four metropolises. The energy storage facilities are deployed around the location with rich renewable generation.
	}
	\label{conceptual}
\end{figure}

In order to provide firm insights into potential technical corrective policy assessment, we start from investigating four possible corrective measures, namely, generation units winterization, interconnection HVDC lines, up-scaled demand response program, and strategic energy storage facility as conceptualized in Fig. \ref{conceptual}.  Generation units winterization only refers to the adapted winterization treatments for electric energy generation units, assuming no winterization is currently applied for any units. Particularly, we reserve the impacts of wide-ranging natural gas scarcity \cite{gas_scarcity} due to failures in the non-winterized natural gas supply chain (see Appendix A.4). We also note that there could have been potential interruption to gas refineries in Texas which might have also contributed to the lack of natural gas supply, as the gas reserve amount within Texas have been low compared to the consumption rate. Assumed as a part of a possible future macro nationwide HVDC network, additional HVDC interconnections, besides two existing HVDC ties, respectively connect from the Western Interconnection to the West zone and from the Eastern Interconnection to the Coast zone, and require necessary transmission lines upgrade around the locations of their converter stations (see Appendix A.8). Up-scaled demand response refers to various incentive programs distributed across ERCOT that require voluntary reduction of electric energy demand. Energy storage refers to the large utility scale storage facilities that absorb the excessive energy during off-peak hours and release it at high power during emergency hours. Here we treat the energy storage as the first-aid measure while taking the other three as the sustained electricity supply measure, especially viewing the generation units winterization as the primary preventive measure. Therefore, we conduct quantitative assessment of all corrective measures from different perspectives in the following analysis.

\subsection{Generator Outage and Corrective Measure Modelling}
To quantitatively assess the impact of generator outage on the severity of the outage event and evaluate the effectiveness of potential corrective measures such as the additional HVDC interconnections, generation units weatherization, up-scaled demand response programs and energy storage, it is necessary to model each of these elements appropriately in our synthetic grid. The detailed modelling method is documented below:

\begin{itemize}
    \item \emph{Generator Outage}: The unit-level generator outage data is retrieved from the ERCOT public report \cite{ercot_generator_outage, ercot_generator_outage2}. As detailed generator information is not available to the public (especially for natural gas, wind and solar generators), we have 606 equivalent generators across the network such that the area-wide aggregated generation capacity and profile can match real ERCOT records. To match the real outage data with synthetic equivalent generator, we pre-process the data by aggregating the total outage capacity for each county and fuel type. The county-level outage data are used to scale the maximum real power output of all corresponding generators in the same county and of the same fuel type in the synthetic network. 
    
    \item \emph{Generator Weatherization}:
    Weatherization is a preventive measure to reduce the impact of extreme weather conditions on the functionality of infrastructures. Different types of generators require different types of weatherization treatments: wind turbines require blade and gearbox heating while gas plants may need anti-frost treatment for facilities. Despite the potentially big difference in weatherization cost and complexity, our focus is more about evaluating the effectiveness of weatherization among different generator types and geographical regions. To this end, we only specify the amount of MW of weatherized generators for each fuel type in each weather zone and try to compare the effectiveness on the reduction of ENS as a result of weatherization. We have also considered the scarcity of natural gas supply before and during the winter storm by approximating the de-rating of the winterized generators caused by gas supply shortage, using disclosed unit de-rating data from ERCOT (see Appendix A.5). When computing the area-level available capacity during counterfactual simulation, we only apply weatherization to generators that were completely out, as the de-rating of the running generators was highly likely caused by lack of gas supply. The weatherized generators are also de-rated based on the extent of de-rating of running gas generators in the same county.
    
    \item \emph{HVDC Ties}: We have included two existing HVDC ties to the Eastern Interconnection in our synthetic network model. These two existing conveter stations are modelled as fixed equivalent loads for DCOPF computation. In each hour, the real tie flow data from ERCOT is assigned as real power demand for the equivalent loads and the sign is determined by the direction of DC tie flow. This is meant to make the DC tie flow exactly the same as real data. For counterfactual studies, we have included two additional HVDC ties, one to the Western Interconnection and another to the Eastern Interconnection. Here we assume a macro HVDC interconnection network will be built across the U.S. following a commonly used design \cite{osborn2016designing}. We have adopted this design in determining the location for the converter stations and upgrade the transmission lines around the converter stations (see Appendix A.8). As the additional HVDC ties are introduced to be counterfactual corrective measures to supply any extra power during the event, their actual output power needs to be adjusted according to the ERCOT system demand. Hence, the new DC ties are modelled differently as equivalent generators with a cost function representing the hypothetical cost to buy power from neighboring states. The magnitude and direction of their tie flow is determined by OPF result.
    \item \emph{Demand Response}: 
    Demand response refers to various incentive programs that encourage voluntary reduction of electric energy demand during peak or emergency hours. Although most demand response programs in ERCOT are still in development at the current stage and mostly target large industrial, commercial or aggregated customers, it has the potential to scale up quickly and provide a valuable demand side resource during similar energy emergencies. In our simulation, the effect of a demand response program is modelled as a form of voluntary load reduction with prior agreement between load serving entities and customers which does not count to ENS. During energy emergencies, load resources from demand response programs are the first to be committed before all of the other more costly steps. Hence, we also prioritize available demand response capacity over storage and forced load shedding. When the system reserve is below 2300 MW which corresponds to ERCOT EEA level 1, the loads across the synthetic network are reduced up to the maximum allowed demand response capacity.
    \item \emph{Energy Storage}:
    The purpose of energy storage is to absorb excessive energy from renewable sources during off-peak hours and release stored energy during peak and emergency hours. Unlike other types of corrective measures that are more suitable in reducing the overall severity of the outage by providing sustained additional power supply, the advantage of energy storage lies in its ability to provide relatively large power output during a short period, which can be used to bridge through "most difficult" hours. To emphasize this, we assume all storage capacities are fully charged prior to the event and commit them during the hours when the level of forced load shed is around the highest. The contribution of energy storage is reflected in the reduction of peak load shed capacity.
\end{itemize}

\begin{figure}[ht]
	\centering
	\includegraphics[width=1\textwidth]{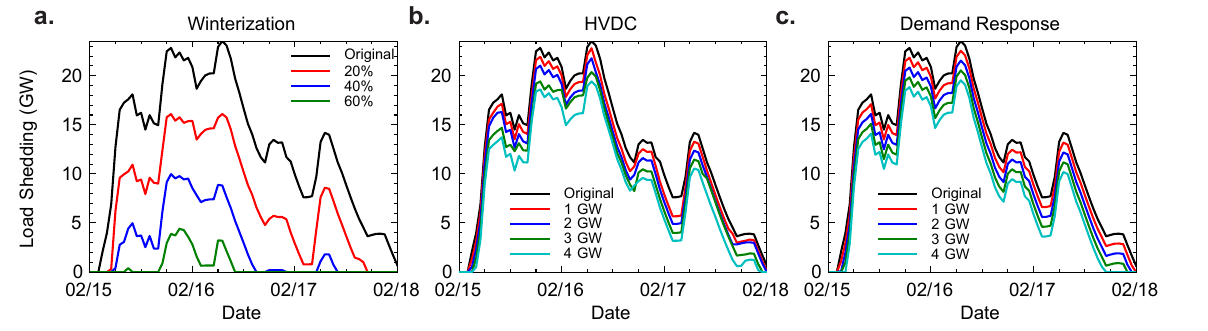}
	\caption{\textbf{Quantitative assessment of each single sustained supply measure in terms of load shedding (GW).} \textbf{a}, The impacts of additional facility winterization on load shedding with different percentages from $20\%$ to $60\%$. These percentages correspond to the winterization for generation capacity of 21.7 GW, 43.4 GW and 65.1 GW. \textbf{b}, The impact of additional HVDC lines on load shedding with the capacity ranging from 1 to 4 GW. \textbf{c}, The impact of up-scaled demand response program on load shedding with the capacity ranging from 1 to 4 GW. Note that winterized generator units and demand response programs are deployed across ERCOT without priority in any specific area. Here the original curve refers to the load shedding in the event reproduction with no corrective measures.
	}
	\label{single-measure}
\end{figure}

Taking these corrective measure settings into account, we first quantify the impacts of each single sustained electricity supply measure of distinct extents in load shedding (Fig. \ref{single-measure}). We find that 60\% generation units winterization can effectively reduce the Energy-not-Served from 929.6 GWh to 40.8 GWh (Fig. \ref{single-measure}-a), and about 80\% generation units winterization can prevent the blackout entirely, where we reserve the impacts of non-winterized natural gas supply chain. We also find that HVDC lines and up-scaled demand response of equal capacity have similar but different effectiveness on mitigating the electricity scarcity (Fig. \ref{single-measure}-b,c), which respectively reduces the Energy-not-Served by $64.1$ GWh and $67.5$ GWh per 1 GW capacity (see Appendix A.6 for more details). 
Since energy system winterization is the most straightforward solution against extreme frigid weather, we attach additional importance to prioritizing the winterization of generation units of specific source types in different regions. We perform the quantitative assessment of the effectiveness of facility winterization by source and region on the electricity scarcity mitigation. The results in Table \ref{tab-1} indicate the distinct performance of per-GW generation units winterization (see Appendix A.7 for more details). Based on this, we suggest the priority of winterization for the disabled nuclear generation units located in the South Central, natural gas generation units across ERCOT, coal generation units in the Coast and wind generation units in the West.

\begin{table}[h]
\centering
\begin{tabular}{lccccccccc}
\hline
Source Type & Far West&West&North&East&Coast&North Central&South Central&South\\
\hline
Nuclear & \NA & \NA & \NA & \NA & \NA & \NA & 38,205 & \NA   \\
 
Natural Gas & $37,058$ & $12,338$ & $18,993$ & $16,908$ & $25,129$ & $29,236$ & $18,905$ & $23,279$ \\
 
Coal & \NA & \NA & \NA & $\phantom{00,}266$ & $18,719$ & $\phantom{0}1,967$ & $\phantom{00,}266$ & $\phantom{0}5,768$  \\
 
Wind  & $\phantom{0}6,865$ & $15,178$ & $\phantom{0}4,479$ & $\phantom{0}5,188$ & $\phantom{00,}795$ & \NA & \NA & \NA  \\
\hline
\end{tabular}
\caption{\textbf{Quantitative assessment of generation units winterization by source and region in terms of Energy-not-Served reduction (MWh)}. Each entry in the table shows the resulting reduction of Energy-not-Served (in MWh) for 1 GW generation capacity winterization of each source type in each weather zone. A crossed-out entry means that the lack of  associated generation outage data renders the evaluation of this certain winterization non-applicable.}
\label{tab-1}
\end{table}

\begin{figure}[h]
	\centering
	\includegraphics[width=1\textwidth]{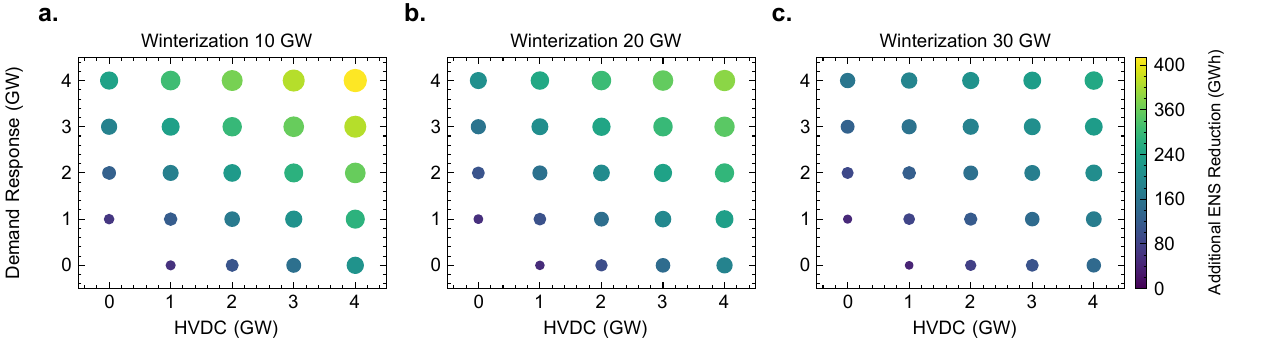}
	\caption{\textbf{Additional Energy-not-Served reduction (GWh) by HVDC and demand response given different winterization portfolios.} \textbf{a, b, c,} respectively show the additional Energy-not-Served reduction contributed by HVDC and demand response that is represented by both bubble size and color, given different total winterized generation capacity. Here the winterization portfolios in three cases are determined based on the results in Table \ref{tab-2}. The Energy-not-Served reduction contributed by winterization portfolio alone are respectively 266.2 GWh, 467.0 GWh and 628.9 GWh for the case of 10 GW, 20 GW and 30 GW winterized capacity.
	}
	\label{portfolio}
\end{figure}

\begin{figure}[h]
	\centering
	\includegraphics[width=0.4\textwidth]{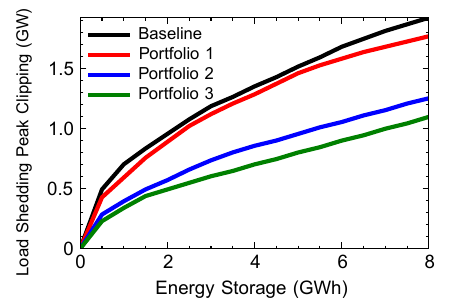}
	\caption{\textbf{Load shedding peak clipping of energy storage facility.} The baseline refers to the model without other corrective measures, while the portfolio 1, 2 and 3 respectively consist of one of three winterization portfolios in Table \ref{tab-2} together with HVDC and demand response of 2 GW capacity.
	}
	\label{energy-storage}
\end{figure}

\begin{table}[!h]
\centering
\begin{subtable}[t]{0.8\textwidth}
\caption*{Allocation of Winterized Capacity (MW) in Portfolio 1}
\begin{tabular}{lrrrrrrrrr}
\hline
Weather Zone & Far West & West & North & East & Coast & North Central & South Central & South \\ \hline
Natural Gas  & 2000     & 0    & 0     & 0    & 2000  & 2000          & 0             & 2000  \\ 
Coal         & 0        & 0    & 0     & 0    & 1000  & 0             & 0             & 0     \\ 
Wind         & 0        & 0    & 0     & 0    & 0     & 0             & 0             & 0     \\ 
Nuclear      & 0        & 0    & 0     & 0    & 0     & 0             & 1000          & 0     \\ \hline
\end{tabular}
\end{subtable}
\begin{subtable}[t]{0.8\textwidth}
\caption*{Allocation of Winterized Capacity (MW) in Portfolio 2}
\begin{tabular}{lrrrrrrrrr}
\hline
Weather Zone & Far West & West & North & East & Coast & North Central & South Central & South \\ \hline
Natural Gas  & 3500     & 500  & 1000  & 1000 & 2500  & 3000          & 1000          & 2500  \\ 
Coal         & 0        & 0    & 0     & 0    & 1000  & 0             & 0             & 500   \\ 
Wind         & 500      & 1000 & 500   & 500  & 0     & 0             & 0             & 0     \\ 
Nuclear      & 0        & 0    & 0     & 0    & 0     & 0             & 1000          & 0     \\ \hline
\end{tabular}
\end{subtable}

\begin{subtable}[t]{0.8\textwidth}
\caption*{Allocation of Winterized Capacity (MW) in Portfolio 3}
\begin{tabular}{lrrrrrrrrr}
\hline
Weather Zone & Far West & West & North & East & Coast & North Central & South Central & South \\ \hline
Natural Gas  & 3500     & 500  & 1000  & 1000 & 2500  & 3000          & 1000          & 2500  \\ 
Coal         & 0        & 0    & 0     & 0    & 1000  & 0             & 0             & 500   \\ 
Wind         & 500      & 1000 & 500   & 500  & 0     & 0             & 0             & 0     \\ 
Nuclear      & 0        & 0    & 0     & 0    & 0     & 0             & 1000          & 0     \\ \hline
\end{tabular}
\end{subtable}
\caption{\textbf{Selective generation units winterization in portfolio 1, 2 and 3}}
\label{tab-2}
\end{table}
Given the quantitative assessment of single sustained electricity supply measure, we investigate the performance of several sustained electricity supply portfolios and assess the first-aid capability of energy storage on the basis of sustained sources. 
We have selected three appropriate winterization portfolios based on the foregoing priority analysis, as shown in Table \ref{tab-2} to provide a quantitative assessment in terms of additional Energy-not-Served reduction contributed by both HVDC and demand response (Fig. \ref{portfolio}). First, we find that the performance of HVDC and demand response are slightly different but almost equivalent under certain cases of winterization portfolio. Second, we find the per-GW performance of HVDC and demand response decreases as the winterization capacity increases.
For first-aid outage mitigation at the load shedding peak hour, we focus on load shedding peak clipping by the strategic energy storage facilities on top of sustained corrective measure portfolios, each refers to one of the three winterization portfolios in Table \ref{tab-2} together with HVDC and demand response of 2 GW. We find that the performance of per-GWh capacity reduces as the total energy storage capacity increases, or along with increasingly sufficient sustained supply corrective measures (Fig. \ref{energy-storage}). 
To summarize the key findings obtained in the foregoing quantitative analysis, we find the strong disparity of generation units winterization of various source types in multiple regions, and the interdependence of per-unit performance of corrective measures, based on the quantitative assessment of certain corrective portfolios.

\section{Concluding Remarks}

We develop an open-source and extendable synthetic electric grid model that could serve as a platform for broader energy research community to quantitatively assess the severity and corrective measures of the 2021 Texas power outage.  Simulation results based on this open synthetic model are shown to have captured key characteristics of the real-world event, demonstrating the model fidelity and uncovering the key regional disparity of load shedding. The quantitative assessment of the corrective measures and portfolios has indicated the strong disparity of winterization effectiveness among generation units of various types in multiple regions and the interdependence of per-unit performance among corrective measures. It can immediately inform policy makers of the priority of generation units winterization, the quantitative assessment of certain portfolios on mitigating the blackout and the necessity of launching systematic investigations on the combined effects of corrective measures, which can potentially be generalized for other regions around the world which are experiencing the dual challenge of energy portfolio transition and extreme weather conditions.

This open-source, cross-domain, data-driven approach to analyzing a real-world power grid during extreme events provides a fresh perspective to allow broader climate and energy research communities to have high fidelity characterization of what  happened and what could have been corrected in  large power grids, as energy systems around the world go through profound transformation. 
The design of this open-source synthetic grid can be easily reconfigured, making it very convenient to conduct further analysis such as HVDC interconnection designs, 2030 projected profiles, renewable and storage scenarios and integrated simulation along with the Western and Eastern Interconnection. 
The transparency and extendability of the synthetic grid model  will contribute to a data-driven technology and policy assessment of the energy system transformation with respect to climate change and extreme weather events.

This model and analysis is a first step towards an open-domain cross-displinary approach to fully understand the impact of severe weather on critical infrastructure systems such as the 2021 Texas power outage. Building upon this open-source model, we hope future research to address several key challenges in assessing the severity and causes of the 2021 Texas blackout. 
First, we notice the wide-ranging natural gas generation capacity outage and de-rating are not simply due to the freezing temperature, but also to natural gas scarcity and interruption in the supply chain. It is particularly important to estimate and predict the impacts of interdependence between two energy infrastructure systems on the overall energy system reliability and energy market stability on both sides under extreme weather conditions. Second, we notice the price-ceiling-hitting whole-sale electricity price \cite{high_price} at \$9,000 per MW, that lasted for three days ending on February 18. Its quantified impacts on generation dispatch and load restoration still remain unknown. More investigation is necessary for demonstrating and developing a benign power market mechanism that can encourage improving the reliability and resiliency of power grids. Third, reproducing the reported frequency event at the beginning of blackout event based on reliable dynamic parameters will provide more fruitful insights into what, how and why it happened. Last but not least, the interdependence of per-unit performance among corrective measures emphasizes the necessity of systemic assessment on the combined cost-effectiveness of technical solution bundles.

\section{Data and Code Availability}

The open-source synthetic grid model and the corresponding dataset used for blackout  simulation are publicly available on Github\cite{git}. A tutorial and examples for running the code using provided dataset are also available in the Github repository.

\Urlmuskip=0mu plus 1mu\relax  
\bibliography{ref}




\clearpage

\section*{Appendices}

\subsection*{A. 1: Remarks on the gap between actual online and estimated generation capacity}
There are two periods that have apparent gap between actual and estimated generation capacity as shown in Fig. 1. The first period is before 12 a.m. February 15, when the EEA 3 has not been launched. The reason for this mismatch is that the estimated generation capacity includes all non-outaged generation that is not necessarily online, and it is normal that generation units are offline according to the dispatch or scheduled seasonal maintenance. Therefore, the estimated generation capacity is higher than the actual online generation capacity. The second period is after 12 p.m. February 16, when the gap becomes increasingly large. We suspect it could involve several causes, of which the main one may be attributed to the generators whose de-rating and outage are not disclosed in the public report \cite{ercot_generator_outage2}. Besides, the good match between 12 a.m. February 15 and 12 p.m. February 16 is because all available generation must be required online under such emergency conditions. Since we focus on the blackout event period, it is reasonable to use the estimated generation capacity instead of the actual online capacity to achieve the most accurate granularity.

\subsection*{A. 2: Remarks on the mismatch between actual and simulated load shedding}
Here we separate the period from February 15 to February 18 into three parts, namely the load shedding rising stage from 0 a.m. to 8 p.m. February 15, the load shedding stable stage from 8 a.m. February 15 to 12 p.m. February 16 and the load restoration stage from 1 p.m. February 16 to 12 a.m. February 18. During the load shedding rising stage, actual and simulated Energy-not-Served are respectively 277,625 MWh and 274,316 MWh, of which the relative error is -1.19\%. During the load shedding stable stage, actual and simulated Energy-not-Served are respectively 292,778 MWh and 331,210 MWh, of which the relative error is 13.12\%. During the load restoration stage, actual and simulated Energy-not-Served are respectively 428,388 MWh and 324,115 MWh, of which the relative error is -24.34\%. The mismatch between actual and simulated total Energy-not-Served is mainly derived from the later two stages. The gap during the load shedding stable stage is mostly due to multiple factors such as system topology, system congestion pattern and load shedding strategy (shown in Methods). However, to our best ability, this is the closest result we can obtain based on only publicly available materials, in which limitations include the low geographical resolution of generation, demand and outage data as well as lack of detailed load shedding protocols or event logs in ERCOT internal documentations. The significant mismatch during the load restoration stage shows that the actual load restoration is slower than the simulated one, which may reasonably attribute to the requirements by system transient stability, unknown load regulation, and unreported technical load restoration issues that nevertheless are beyond this paper's scope and need more attention for future research.

\subsection*{A. 3: Remarks on the benchmark of load shedding allocation}
Although we have acquired county-level outage data from PowerOutage.com \cite{poweroutage} that show the number of customers with and without electricity during the Texas windstorm event, those data are not appropriate for simulation purposes as they do not provide actual online and offline capacity (in kW/MW). In our simulation, the loads are allocated and scaled based on their rated MW capacity and total ERCOT historical hourly load in each weather zone. While it would be intuitive to draw a direct relationship between the number of disconnected customer and the total capacity of those customers, we unfortunately do not have the necessary data with high enough resolution to do so as the load data in our synthetic network is aggregated and represents the total capacity of entire towns which include residential, commerical and industrial load altogether.

\subsection*{A. 4: Relationship between natural gas generation derating and gas supply scarcity}
Our hypothesis is that all natural gas generation derating is derived from the gas supply scarcity and the remaining full outage is derived from equipment failures at power plants. The Texas blackout event review \cite{ercot_review} includes some information related to the generation outages, which documented that the  cumulative generation capacity forced out throughout the event is 46,249 MW, cumulative number of generators outage throughout the event is 356 and cumulative gas generation de-rated due to supply issues is 9,323 MW. Here the "cumulative capacity" includes all units that have failed at some point, regardless of whether it comes back later during the event, as defined as in the 2011 Texas winter event report \cite{winter_2011}. To verify the hyphothesis, we calculate the cumulative generation capacity forced out and cumulative number of generators outage based on the generation outage report \cite{ercot_generator_outage2}, which are respectively 47,946 MW and 316. The relative error between estimated and reported values are 3.67\% and 11.23\%, which validates the correctness of the calculate method. The mismatch between real and reported cumulative number of generator outages is in line with our expectation because about 10\% of generation plants deny to release of the outage information to the public \cite{ercot_generator_outage2}. In the same way, we calculate the cumulative generation derating that equals to 10,608 MW, which has 13.78\% relative error compared to the reported 9,323 MW. To simplify the problem, we confirm that the proposed hypothesis roughly matches the reality which indicates that all natural gas generation derating are cause by gas supply scarcity.

\subsection*{A. 5: Incorporating gas supply scarcity into natural gas thermal generation winterization}
As shown in Appendix A.1, we are able to roughly separate the outage/de-rating of generators that are caused by lack of gas supply and those that are caused by un-winterized power plant facilities. With this information, we are able to incorporate gas supply interruption in our counterfactual simulations that involve the hypothetical weatherization of natural gas generators. We only apply weatherization treatments to generators that are completely out-of-service as shown in the ERCOT unit outage data. For each of those generators, we calculate the amount of de-rating of  other in-service but de-rated generators in the same county by assuming that the availability of gas supply is roughly the same across gas generators in the same county. Hence, even when those completely out generators are in service as a result of weatherization, they still cannot run at their maximum capacity due to the lack of gas supply. In counterfactual simulation, the de-rating caused by gas supply is done by giving additional natural gas generators a de-rating multiplier that is determined by the level of de-rating of its neighbouring generators in the same county.

\subsection*{A. 6: Remarks on the performance of HVDC and demand response}
In Fig. 4b and 4c, the effect of HVDC and demand response on forced load shedding is different even for the same total capacity. This difference is mainly caused by the difference in modelling these two types of corrective measures. For the modelling of additional HVDC, we have added two converter stations into the synthetic network: one in the city of Roscoe in the West zone representing a DC tie to California, one in the city of Bryan in the Coast zone representing a DC tie to Florida. Transmitting power from these two converter station to the rest of the grid is limited by network congestion, which can result in remote loads not getting power from HVDC interconnections and thus they must be shed even when there is available power supply from the HVDC ties. In contrast, for the modelling of demand response, we assume the additional load-capacity is split across the entire network, which is less likely to cause congestion than concentrated high-capacity energy supply like the HVDC converter stations. Moreover, since the system demand is still reduced after introducing demand response (but on a voluntary basis), the lower demand also alleviates the congestion pattern in the network which leads to more effective energy use.

\subsection*{A. 7: Remarks on the performance of winterization of various sources in multiple regions}
Appendix Table A.1 has shown that for the same amount of MW of weatherization, the difference in effectiveness of different fuel types can be very different. This difference is mainly caused by the severity of outage and derating during the event across generator types. Some type of generators, such as natural gas, are more vulnerable to cold weather than others and have suffered much higher levels of outage and derating, thus weatherization treatment would be more effective. In contrast, coal and nuclear generators are much less affected by the winter storm (probably due to their larger size), thus additional weatherization treatment won't improve the situation much. In short, the more severe a generator is affected, the more effective will its weatherization be. Moreover, even with complete weatherization, the actual capacity of wind turbines is affected by the wind strength. For example, the wind strength between 9 a.m. February 15 and 8 p.m. February 16 is weak, hence during this time period wind turbines cannot provide much power even if they are fully weatherized.

\subsection*{A. 8: Transmission line upgrade around the interconnection nodes of the HVDC lines}
In our counterfactual case studies we have added two additional HVDC converter stations that represent ties to California and Florida to the original synthetic grid network. Each station is capable of transmitting at a maximum 2000 MW of real power into the Texas network. This additional power injection will significantly overload the AC transmission grids near the point of common coupling as their specifications are not designed to handle the additional power flow. To accommodate the new resources, we have up-scaled the thermal limit of AC transmission lines in the neighbouring region to avoid line overflowing. We have designed a iterative algorithm to upgrade the AC transmission line appropriately without disrupting the congestion pattern in the original network.

We first model the topology of a transmission network as an un-directed graph, where each bus is represented as a vertex and each branch is represented as an edge. Let $B$ denote the set of all buses (vertices) in the graph; let $L = {L_{ij}} \; \forall i,j \in B$ denote the set of all branches (edges). The \emph{distance} between two vertices $i$ and $j$ is denoted as $D_{ij}$. The real power flow in a line $L_{ij}$ is $P_{ij}$. For each transmission there exists a thermal rating, $P^{max}$, that dictates the maximum allowed power flow along the line. The thermal limit is used in OPF formulation to ensure that the power flow of all lines in the network is lower than the thermal limit, $P_{ij} <= P_{ij}^{max} \; \forall i,j \in B$. We define the line Load Factor $\rho_{ij} = P_{ij} / P_{ij}^{max}$ as the ratio between the line flow to its thermal limit. With the addition of a new power source such as an HVDC converter, the line flow pattern across the network will change to reflect the new power flow solution. Most likely, the power flow in the lines that are close to the converter station will increase drastically as the capacity of HVDC lines is usually much larger than the power absorption of local loads in the region and will cause lines to overflow. To upgrade those lines appropriately, we take an iterative approach to ensure that the load factor $\rho$ of lines in nearby regions remain unchanged before and after adding the new HVDC tie with a designated max capacity. The detailed algorithm is presented as follow:

\begin{algorithm}
    \caption{AC Transmission Network Upgrade for Additional HVDC Links}
	\label{algo:2s}
    \begin{algorithmic}
    \STATE Define the maximum capacity of new HVDC tie $P_{dc}$ and increment stepsize $\Delta P_{dc}$
    \STATE Define the maximum distance $N$ within which the lines will be upgraded
    \STATE Select a heavily loaded profile snapshot as baseline
    \STATE Select a bus $o \in B$ in the network as the location for new DC converter station
    \STATE Find the sets of buses $[B^1, B^2, .., B^N]$ such that $D_{oj} == n, \forall j \in B^n$ through recursive graph tracing
    \STATE Find the sets of all lines $[L^1, L^2, .., L^N]$ that connects the buses of neighbouring distance: $L^n = \{L_{ij} \; | i \in B^{n-1}, j \in B^{n}\}$
    \STATE Run DCOPF to determine initial line flows $P_{ij}^{base} \; \forall i,j \in B$
    \STATE Calculate load factor for lines that require upgrading, $\rho_{ij}^{base} \; \forall i,j \in [L^1, L^2, .., L^N]$
    \STATE Initialize the capacity of new DC tie, $P^{dc} = \Delta P_{dc}$
    \WHILE{$P^{dc} < P_{max}^{dc}$}
    \STATE Set the power injection at bus $o$, $P_o = P_{dc}$
    \STATE Solve DCOPF to obtain new line flows with additional HVDC injection $P_{ij} \; \forall i,j \in B$
    \FOR{n = [1, 2, .., N]}
    \STATE Upgrade thermal rating of lines $P_{ij}^{max} = \max (P_{ij} / \rho_{ij}^{base},\:P_{ij}^{max}) \; \forall i,j \in L^n$
    \ENDFOR
    \STATE Increment $P_{dc} = P_{dc} + \Delta P_{dc}$ iteratively to ensure convergence
    \ENDWHILE
    \end{algorithmic}
\end{algorithm}

\subsection*{A. 9: Remarks on the simulation environment for this open-source model}

Although the model itself adopts the Matpower format for storing the bus, generator and topology information, it its not exclusive to Matpower which is based on MATLAB. We chose Matpower to perform our studies for this paper but similar works can be done using other platforms when MATLAB is not available. For example, there exists several free Python-based software that can also run analysis directly on the Matpower case format such as Pandapower \cite{PandaPower} and REISE \cite{reise}. In fact, some of our preliminary works on the creation and calibration of the synthetic model were carried out using REISE without MATLAB.

\end{document}